\documentclass[fleqn,twoside]{article}
\usepackage{espcrc2}

\usepackage{graphicx}
\usepackage[figuresright]{rotating}
\input{epsf}
\newcommand{\beq}{\begin{equation}}
\newcommand{\eeq}{\end{equation}}
\newcommand{\bea}{\begin{eqnarray}}
\newcommand{\eea}{\end{eqnarray}}
\newcommand{\bmp}{\begin{minipage}}
\newcommand{\emp}{\end{minipage}}
\newcommand{\D}{\displaystyle}
\newcommand{\tx}{\textstyle}
\newcommand{\s}{\scriptstyle}

\newcommand{\bS}{{\bf S}}

\newcommand{\vev}[1]{\Big\langle #1 \Big\rangle}

\newcommand{\AmS}{{\protect\the\textfont2
  A\kern-.1667em\lower.5ex\hbox{M}\kern-.125emS}}

\title{Loop inequalities and confinement}

\author{E.T. Tomboulis\address{Department of Physics, UCLA, Los Angeles, 
CA 90095-1547, USA}}
       
\begin{document}

\begin{abstract}
We consider correlation inequalities that follow from the 
well-known loop equations of LGT, and their analogues in spin systems. 
They provide a  
way of bounding long range by 
short or intermediate range correlations. 
In several cases the method easily reproduces   
results that otherwise require considerable effort 
to obtain. 
In particular, in the case of the 2-dimensional  O(N) spin model, 
where large N analytical results are available, the 
absence of a phase transition and the exponential decay of 
correlations for all $\beta$ is easily demonstrated.   
We report on the possible application of this 
technique to the analogous 4-dimensional problem of area law for 
the Wilson loop in LGT at large $\beta$.

\vspace{1pc}
\end{abstract}

\maketitle
\vspace{-0.3cm}
\section{Loop equations in LGT} 
The loop equations of gauge theory, i.e. the SD equations for 
Wilson loops, are well-known both on the lattice and, at least 
formally, also in the continuum \cite{FEMM}. On the lattice:    
\beq \bmp{1.3cm}
\epsfysize=1.3cm\epsfbox{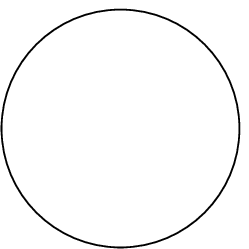}\emp = \beta\,\sum_{\hat{\mu}} \;
\bmp{1.5cm} \epsfysize=1.5cm\epsfbox{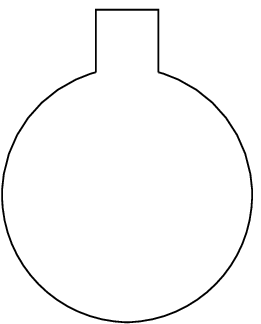}\emp\!\! 
-\, \beta\,\sum_{\hat{\mu}} \; \bmp{1.6cm} \epsfysize=1.6cm\epsfbox{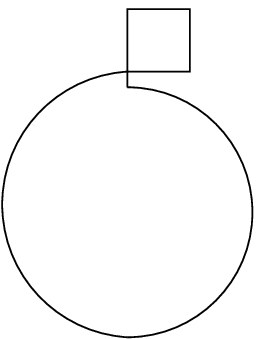}
\emp \label{L1}  
\eeq
Here, and in the following, circles denote general but 
{\it non-self-intersecting} loops. 
(For self-intersecting loops, the l.h.s. contains additional terms 
involving multiple-loop expectations generating the infinite sequence 
of coupled SD loop equations. For a simple loop, however, one has the 
closed equation (\ref{L1}).) The deformations on the r.h.s. 
involve one plaquette protruding in direction $\mu$. Also:  
${\D \beta=1/Ng^2 \qquad \mbox{for} \; U(N)}$, 
${\D \beta=N/[(N^2-1)g^2\,] \;\:\mbox{for}\;SU(N)}$. 
\vspace{-.2cm}
\section{Basic idea}
In strong coupling expansion for $U(N)$ one finds for the `curly' 
deformation term 
\beq
 \sum_{\hat{\mu}} \; \bmp{1.5cm} \epsfysize=1.5cm\epsfbox{cl.eps}\emp 
\Bigg/ \bmp{1.4cm}
\epsfysize=1.4cm\epsfbox{l.eps}\emp =(2d-3)\beta + \cdots 
\eeq
so the loop equation (\ref{L1}) gives 
\beq 
\bmp{1.5cm}
\epsfysize=1.3cm\epsfbox{l.eps}\emp = {\beta\over 1 + \beta^2(2d-3) + \cdots} 
\sum_{\hat{\mu}} \;\bmp{1.3cm}
\epsfysize=1.5cm\epsfbox{pl.eps}\emp\;\label{L2}
\eeq 
Now one may iterate (\ref{L2}) any number of times up to the number of 
plaquettes in the minimal loop area $|A|$, and obtain \cite{TUW}:   
\beq 
\bmp{1.5cm}
\epsfysize=1.5cm\epsfbox{l.eps}\emp \leq \lambda^{|A|}\;\bmp{1.3cm} 
\begin{center} 
{Max\\ all loops}\end{center}\emp \;
\left|\bmp{1.5cm}
\epsfysize=1.5cm\epsfbox{l.eps}\emp \right| 
\eeq 
i.e. area law 
\beq 
\bmp{1.5cm}
\epsfysize=1.5cm\epsfbox{l.eps}\emp \leq \mbox{Const.} \;
\exp\left[-(\ln{1\over\lambda})\,|A|\right] 
\eeq
{\it provided} \quad ${\D \lambda \equiv {2(d-1)\,\beta\over 
1 + \beta^2(2d-3)} < 1}$, 
i.e. for \quad $\D \beta < {1\over (2d-3)}$. 
Note that this estimate is comparable to estimates 
of the convergence radius of the strong coupling expansion 
obtained after rather more involved arguments.

Abstracting from the strong coupling case suggests the 
following general approach \cite{TUW}. Assume that the `curly' term 
satisfies 
\beq 
\sum_{\hat{\mu}} \; \bmp{1.5cm} \epsfysize=1.5cm\epsfbox{cl.eps}\emp 
\geq  \;\alpha(\beta) \;\bmp{1.3cm}
\epsfysize=1.3cm\epsfbox{l.eps}\emp 
\eeq  
for some function $\D \alpha(\beta)$ (for sufficiently large loops). 
Then the loop equation (\ref{L1}) gives 
\beq 
\bmp{1.3cm}
\epsfysize=1.3cm\epsfbox{l.eps}\emp \leq {\beta\over 1 + \beta \alpha(\beta)} 
\sum_{\hat{\mu}} \;\bmp{1.4cm}
\epsfysize=1.4cm\epsfbox{pl.eps}\emp\; \label{Lineq} 
\eeq 
and area law follows by the above iteration argument, 
previously applied to (\ref{L2}), provided  
\beq  
{\beta2(d-1)\over 1+\beta \alpha(\beta)} < 1\label{alphab} 
\eeq 


Attempts to extract area-law from the continuum loop equations  
have not been particularly successful. Very little seems to have been 
done with the lattice version. 
Before proceeding to our main case of interest, i.e $4$-dimensional 
LGT at large $\beta$, we illustrate this general approach in 
some simpler examples. 

\section{U(N) LGT in 2 dimensions} 
In this case we have the equality: 
\bea 
\sum_{\hat{\mu}} \;\bmp{1.5cm}
\epsfysize=1.5cm\epsfbox{cl.eps}\emp & = & \left[\, 
\bmp{0.4cm}
\epsfysize=0.3cm\epsfbox{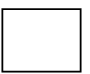}\emp + 
{\bmp{0.5cm}
\epsfysize=0.5cm\epsfbox{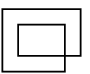}\emp  \over 
\bmp{0.4cm}
\epsfysize=0.3cm\epsfbox{p1.eps}\emp } \;\,\right] \; 
\bmp{1.3cm}
\epsfysize=1.3cm\epsfbox{l.eps}\emp \nonumber \\
& \equiv & \alpha(\beta) \;\, \bmp{1.3cm}
\epsfysize=1.3cm\epsfbox{l.eps}\emp \;.
\eea
Substituting in the loop equation gives (\ref{Lineq}) as an equality,  
and, from (\ref{alphab}), area law follows if ${\D \alpha(\beta) 
> 2-{1\over \beta}}$.   
Now, explicit computation in the large $N$ limit gives: 
\bea 
\alpha(\beta)&=&\beta \; , \quad \beta<{1\over2} \qquad \nonumber\\
  &=&2-{1\over \beta} + {1\over 4\beta(4\beta-1)}\,, \quad \beta>{1\over2}\,. 
\nonumber 
\eea
So the above bound on $\alpha(\beta)$ is satisfied for all 
$0 \leq \beta <\infty$. 

\section{O(N) spin model in 2 dimensions} 
In the case of the spin model the analog of loop equation (\ref{L1}) is  
\bea 
\vev{\bS_n\cdot\bS_m} & = & \beta\,\sum_{\mu=\pm1}^{\pm2} 
\vev{\bS_{n+\hat{\mu}}\cdot \bS_m} \label{S1}\\  
 & & \quad -\, \; \beta\,\sum_{\mu=\pm1}^{\pm2} 
\vev{\bS_{n+\hat{\mu}}\cdot \bS_n \bS_n\cdot\bS_m},\nonumber 
\eea 
where $\bS_n$ denotes an $N$-component unit length spin at site $n$. 
Note that the second term on the r.h.s corresponds to the `curly' term 
and comes with a minus sign. 
We work in the large $N$ limit. To leading order in $1/N$:
\beq
\vev{\bS_{n+\hat{\mu}}\cdot \bS_n \bS_n\cdot\bS_m} = 
\vev{\bS_{n+\hat{\mu}}\cdot \bS_n}\vev{\bS_n\cdot\bS_m} \label{fact} 
\eeq 
So (\ref{S1}) becomes
\beq
\vev{\bS_n\cdot\bS_m} = {\beta\over 1 + \beta\alpha(\beta)}\,
\sum_{\mu=\pm1}^{\pm2} 
\vev{\bS_{n+\hat{\mu}}\cdot \bS_m}.\label{Sineq}
\eeq
This is the analog of (\ref{Lineq}), and the same reasoning implies that   
$\vev{\bS_n\cdot\bS_m}$ decays exponentially as long as:  
\beq
\alpha(\beta)\equiv \sum_{\mu=\pm1}^{\pm2} 
\vev{\bS_{n+\hat{\mu}}\cdot \bS_n} > 4 - {1\over \beta} \,.\label{Sineq1}
\eeq 
Now it is a well-known result \cite{BK} that to leading ${1\over N}$ order
\beq
\alpha(\beta) = 4-{1\over\beta} + 2\,e^{-2\pi\beta} + \cdots\,,\qquad 
\beta\to \infty .\label{Scond} 
\eeq  
Hence condition (\ref{Sineq1}) is fulfilled all the way to $\beta\to 
\infty$ proving the exponential decay of correlations and absence of 
a phase transition in the model. Note that this happens by virtue of  
the nonperturbative (exponential) term in 
(\ref{Scond}) signaling the presence of a spin condensate \cite{NSVZ}. 
The next to leading $1/N$ order is rather more 
involved \cite{NSVZ} with technical complications of the sort 
one encounters in the 4-dim gauge theory case.     

\section{4-dim. SU(N) gauge theory at large ${\bf \beta}$} 
We begin by trivially rewriting the curly loop in the loop equation 
(\ref{L1}) as  
\beq 
\bmp{1.5cm} \epsfysize=1.5cm\epsfbox{cl.eps}\emp =   
(1-\kappa)\; 
\bmp{1.4cm} \epsfysize=1.5cm\epsfbox{cl.eps}\emp 
+\;\kappa \;\;  
\bmp{1.5cm} \epsfysize=1.5cm\epsfbox{cl.eps}\emp
\label{L3} 
\eeq  
with $\kappa$ an arbitrary parameter to be adjusted later 
for optimizing bounds.

Inserting (\ref{L3}) in (\ref{L1}), and by a series of manipulations 
involving adding and subtracting appropriately chosen loop terms on 
the r.h.s. and use of reflection positivity, one may derive from 
(\ref{L1}) the inequality:  
\bea 
\beta\,2(d-1)\; 
\bmp{1.3cm}
\epsfysize=1.3cm\epsfbox{l.eps}\emp \;
\Bigg[\,1 & + & {1-r(\kappa) \over \beta\,2(d-1)} + \kappa \nonumber \\
- \;\;[1 + c^2 \; \bmp{0.3cm}
\epsfysize=0.4cm\epsfbox{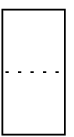}\emp     
-2c \;\bmp{0.2cm}
\epsfysize=0.2cm\epsfbox{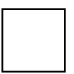}\emp \,]^{1/2}\,\Bigg]      
& \leq & \beta\sum_{\hat{\mu}} 
\bmp{1.4cm}
\epsfysize=1.4cm\epsfbox{pl.eps}\emp \label{L4}
\eea 
where $c=1-\kappa$, and $r$ is a somewhat complicated expression 
involving differences 
of ratios of loops whose explicit form need not be given here.  
$r = O(1)$ but is $\kappa$-dependent. In fact alternative versions 
of this type of inequality with somewhat different forms of $r$ 
may be obtained.   
Writing for the two- and one-plaquette expectations in (\ref{L4}): 
\[\bmp{0.3cm}\epsfysize=0.5cm\epsfbox{tp.eps}\emp = 
1 - \bar{\Sigma}(g),\quad 
 \bmp{0.3cm}\epsfysize=0.3cm\epsfbox{p.eps}\emp = 1- \Sigma(g) \] 
one has:   
\bea 
[1 + c^2 \; \bmp{0.3cm}
\epsfysize=0.5cm\epsfbox{tp.eps}\emp     
-2c \;\bmp{0.3cm}
\epsfysize=0.3cm\epsfbox{p.eps}\emp \,] & = &
\kappa^2\,(1-\bar{\Sigma}) + \kappa\, 2(\bar{\Sigma}-\Sigma) \nonumber \\
  & & +(2\Sigma - \bar{\Sigma}). \label{short}
\eea
Now expanding for small lattice spacing $a$ and using OPE:  
\bea
(2\Sigma - \bar{\Sigma}) = - a^4\,\vev{\,g^2\,F_{\mu\nu}(x)F^{\mu\nu}(x+a)\,} 
+ \ldots & & \nonumber \\
\sim \;\,C_1(a,\mu)\,{\bf 1}  + C_{F^2}(a, \mu)\,\vev{g^2 \bf{F^2}(\mu)} + 
\ldots\qquad  
& & \nonumber
\eea 
with leading singular behavior: 
\[ C_1(a,\mu) \sim a^{-4}\;, \qquad C_{F^2} \sim a^0\;. \] 
We now write the r.h.s. of (\ref{short}) as:   
\beq  
[1 + c^2 \; \bmp{0.3cm}
\epsfysize=0.5cm\epsfbox{tp.eps}\emp     
-2c \;\bmp{0.3cm}
\epsfysize=0.3cm\epsfbox{p.eps}\emp \,]
\equiv [\, {\cal O}(\kappa) - a^4\,{\cal O}_1 \,] \nonumber 
\eeq 
The $\kappa$ independent terms proportional to $a^4$ on the r.h.s. of 
(\ref{short}) are uniquely 
picked out, and arise from the gluon condensate: 
\beq
{\cal O}_1 \sim {\rm Const}\;\Lambda^4 
\sim {\rm Const}\;\mu^4\,e^{-2b_0/g^2(\mu)} \label{cond}
\eeq

After some manipulation the expression multiplying 
the loop on the l.h.s. in (\ref{L4}) can be written as 
\beq 
\beta 2(d-1)\left[\, 1 + {\tx {1\over 2}}\,a^2\,{\cal{O}}_1^{\s 1/2}  
+ f(g^2, \mu, a, \kappa) \,\right]\, , 
\eeq  
where $f$ is given in terms of 
$\Sigma$, $\bar{\Sigma}$ and the other quantities entering in 
$r(\kappa)$. It involves a series in $g^2$ `perturbative' part. 
Note that all $\mu$ dependence, other than in RG invariant combination (as in 
$\Lambda$), arising through the use of OPE must, in principle, cancel 
among the various terms in $f$. 

(\ref{L4}) is optimized by choosing $\kappa$ to maximize $f$.  
Now since the inequality is rigorously valid, $f$ must be 
negative for all $\kappa$ not equal to the maximizing $\kappa_{\rm max}$, 
thus rendering the iteration argument 
inapplicable. Otherwise, 
as it is easily seen, unphysical loop behavior would result. 
For $\kappa_{\rm max}$, the optimum that could be 
achieved is that $f$ vanishes (to within powers of $a\Lambda$).  
In such a case (\ref{L4}) would give:  
\[ 
\bmp{1.2cm}
\epsfysize=1.2cm\epsfbox{l.eps}\emp\ \leq 
{1 \over 2(d-1)\,\left[\,1 + {\tx {1\over 2}}\, a^2\,{\cal{O}}_1^{\s 1/2}
\,\right]} \,\sum_{\hat{\mu}} \; \bmp{1.2cm}
\epsfysize=1.3cm\epsfbox{pl.eps}\emp  \label{L5} 
\] 
which may be iterated as above to give area law with string tension 
$\sim {\rm const}\, \Lambda^2$.

Now $f$ at $\kappa_{\rm max}$ depends rather delicately on 
the structure of $r$ in (\ref{L4}), and is, unfortunately, not 
easily estimated accurately enough to ascertain whether it vanishes. 
This is of course only to be expected. The inequality (\ref{L4}) 
must clearly be very sharp 
in order to obtain the right behavior. As noted there are  
different versions of it with somewhat different forms of $r$ that may 
be explored, as well as possible ways of successive refinement 
of a given form. These are currently under investigation.


\begin{thebibliography}{9}
\bibitem{FEMM} D. Foerster, Phys. Lett. 87B (1979) 87;  
T. Eguchi, Phys. Lett. 87B (1979) 91;  
Y.M. Makeenko and A.A. Migdal, Phys. Lett. 88B (1979) 135;
A.M. Polyakov, Phys. Lett. 82B (1979) 247.    
\bibitem{TUW} E.T. Tomboulis, A. Ukawa, and P. Windey, 
Nucl. Phys. B180 [FS2] (1981) 294.        
\bibitem{BK} T.H. Berlin and M. Kac, Phys. Rev. 86 (1952) 821. 
\bibitem{NSVZ} V.A. Novikov, M.A. Shifman, A.I. Vainshtein and 
V.I. Zakharov, Nucl. Phys. B249 (1985) 445.     
 

\end{thebibliography}
\end{document}